\documentstyle[aps,epsfig]{revtex}
\begin{document}

\centerline{\bf THE STRUCTURE OF THE RIPPLE PHASE IN CHIRAL}
\centerline{\bf AND RACEMIC DMPC MULTIBILAYERS}
\vspace{.4cm}
\centerline{Kheya Sengupta$^1$\footnote{Electronic mail: kheya@rri.ernet.in}, V.A. Raghunathan$^1$\footnote{Electronic mail: varaghu@rri.ernet.in}
 and John Katsaras$^2$\footnote{Electronic mail: KATSARASJ@CP7.AECL.CA}}
\vspace{.5cm}
\centerline{$^1$Raman Research Institute, Bangalore - 560 080, India.}
\centerline{$^2$National Research Council, Steacie Institute of Molecular Sciences, Chalk River Laboratories,\\}
\centerline{Chalk River,  Ontario, K0J 1J0, Canada.\\}

\noindent
{\bf ABSTRACT}\\
\noindent
We present electron density maps of the ripple phase of chiral and racemic
dimyristoylphosphatidylcholine. 
The structures of the two systems are found to be identical within
experimental errors,
thus unambiguously showing that the chirality of the lipid
molecules does not influence the structure of this phase.
\\


\noindent
{\bf I. INTRODUCTION}
\\\\
\indent
Lipid water mixtures show a variety of phases as a function of temperature 
and humidity, including a number of diverse lamellar phases [1-3].  
These
lamellar phases are of interest because they serve as model systems for
biological membranes. Further, phase transitions in these systems is
also a topic of current interest [2].
In the high temperature {\sl L$_\alpha$} phase, which
has the symmetry of a smectic-A liquid crystal, the 
hydrocarbon chains are molten and exhibit no in-plane ordering.
In the low temperature 
{\sl L$_{\beta^{\prime}}$} phase, the chains are in the all-{\it trans} 
conformation and are ordered on a 
two-dimensional lattice 
[1,4]. In addition to these two lamellar 
phases, some phospholipids
also exhibit an intermediate ripple or {\sl P$_{\beta^{\prime}}$} phase, 
characterized by periodic one-dimensional height modulation of the bilayer 
[1-3, 5-7].  In this phase, the conformation of the chains has not yet been 
unambiguously determined, but from x-ray data it is clear that most of the 
hydrocarbon chains are ordered in the same manner as in the 
{\sl L$_{\beta^{\prime}}$} phase.
\\
\indent 
A complete theory of the ripple phase that explains all the experimental 
observations is yet to be formulated. Recently, Lubensky and MacKintosh 
[8,9] have proposed a Landau theory that describes phase transitions 
between the  {\sl L$_\alpha$}, {\sl P$_{\beta^{\prime}}$}  and 
{\sl L$_{\beta^{\prime}}$}
phases in chiral and achiral bilayers. 
In the case of achiral 
bilayers, this model predicts the existence of two distinct symmetric
ripple phases in 
addition to a square lattice phase. When the system is chiral, one of these 
two phases becomes asymmetric [10]. Thus, according to this
phenomenological model, 
achiral bilayers can exhibit only symmetric ripples, while both symmetric 
and asymmetric ripples can occur in chiral bilayers.
\\
\indent
In order to test the predictions put forth by Lubensky and 
MacKintosh [8] concerning the influence of molecular chirality on the 
structure of the ripple phase, Katsaras and Raghunathan [11] carried out 
x-ray diffraction experiments on aligned films of chiral ({\sl l}) and 
racemic ({\sl dl}) dimyristoylphosphatidylcholine (DMPC) bilayers. They 
found that under similar experimental conditions, the diffraction patterns 
from the two systems were practically indistinguishable. Moreover,
both the 
systems were found to have an oblique unit cell, indicating the presence of 
asymmetric ripples.
\\
\indent  
The observation of  asymmetric ripples in the racemate can still be 
reconciled with  the predictions of the Lubensky-MacKintosh model if the 
{\sl d} and {\sl l} enantiomers within each layer phase separate into chiral 
domains. To investigate this possibility, Katsaras et al. performed 
calorimetric measurements on mixtures of {\sl l}-dipalmitoylphosphatidylcholine 
having perdeuterated hydrocarbon chains ({\sl l}-DPPC-d$_{62}$) and 
{\sl d}-DPPC [12]. They found the data consistent with a binary system 
whose components exhibit complete mutual solid solubility. This
suggests that the asymmetric 
ripples seen in racemic DMPC multibilayers are not the result
of the formation of chiral domains due to phase separation.
\\
\indent
The structural features that can be obtained directly from a diffraction
pattern are the following: 
the wavelength of the ripples 
(${\lambda}$),  the average separation between the bilayers 
(${d}$) and  the angle ($\gamma$) between the two axes of the
unit cell. The values of these three parameters 
are found to be comparable in chiral and racemic DMPC bilayers 
under similar experimental conditions [11]. 
In order to check whether the 
detailed shape of the ripples in these two systems are also similar, we
have now calculated the electron density
profiles of the ripple phase of {\sl l} and {\sl dl}-DMPC.  
In this brief report we present the results of these calculations.
All the structural features of the ripple
phase in the two systems are found to be identical within experimental
errors. This confirms the conclusion of ref. [11] that molecular chirality
does not play any significant role in determining the structure of the
ripple phase.\\
\\
\noindent
{\bf II. EXPERIMENT}
\\
\indent
The sample was deposited on a curved glass surface from a concentrated 
lipid/methanol solution to obtain a stack of bilayers 
oriented parallel to the surface. However, the ripple direction does 
not get fixed by this procedure. The result is a sample that is aligned 
along the layer normal but is a powder in the plane of the bilayer. 
The temperature was controlled to within ${\pm }$0.5${^o}$C and the 
relative humidity was 
maintained at 98${\pm}$2${\%}$ . Further experimental details are to be 
found in ref. [11].
\\ \\
\noindent
{\bf III. DATA ANALYSIS}
\\
\indent
We have applied geometrical corrections to the observed intensities relevant 
to our system, taking into account the fact that the sample is a powder 
in two dimensions and oriented in one. We have ignored absorption corrections 
as the thickness of the sample was not accurately known. However, 
by assuming reasonable values of the sample thickness,
 we
have confirmed that
the electron density profiles are  not  
significantly affected by these
corrections [13].
\\ 
\indent 
In order to calculate the electron density map, the phase of each Bragg 
reflection must be 
determined. As the structure possesses a center of symmetry, the phase can
only take values of either 0 or $\pi$.
To determine the phases of the reflections we have adopted the procedure
developed by Sun et. al. [14]. 
The electron density function $\rho$(x,z) is described as the
convolution of a ripple contour function C(x,z) and the transbilayer
electron density profile T$_\psi$(x,z). Here $\hat{x}$ is the direction of the ripple
wavevector and $\hat{z}$ is the direction of the average layer normal.
C(x,z) = $\delta$(z-u(x)), where u(x) describes the ripple profile and is taken to
have the form of a sawtooth with peak-to-peak amplitude A. $\lambda_{1}$  is the
projection of the longer arm of the sawtooth on the x-axis. T$_\psi$(x,z) gives
the electron density at any point (x,z)  along a straight line, which
makes an angle $\psi$ with the z-axis. The electron density in the methylene
region of the bilayer is close to that of water and is taken as zero.
T$_\psi$(x,z) is modeled as consisting of two delta functions with positive
coefficient of magnitude $\rho_{H}$, 
corresponding to the headgroup regions seperated by a
distance L, and a central delta function with negative 
coefficient of magnitude $\rho_{M}$,
corresponding to the methyl region. The six adjustable parameters in the
model are: A, $\lambda_{1}$, $\psi$, $\rho_{H}$/$\rho_{M}$, 
L and a common normalizing factor.
Using this
model for the electron density profile,  
the expected structure factors at the observed (h,k) values are calculated. 
The calculated structure factors are compared with the observed ones and 
a chi-square value is obtained, which is subsequently minimized by varying 
the adjustable parameters in the model. The phase of each of the Bragg 
reflections is obtained from the structure factors calculated from the 
converged model. These calculated phases are combined with the observed 
magnitudes of the structure factors and inverse Fourier transformed 
to get the electron density function.
\\ \\
\noindent
{\bf IV. RESULTS AND DISCUSSION}
\\
\indent
The observed and calculated structure factors ({\bf F$_{o}$} and {\bf F$_{c}$}
respectively) 
for {\sl l} and {\sl dl}-DMPC
at 24{$^o$}C and 98${\%}$ RH is given in Table 1. Almost all the phases 
of the 
corresponding reflections in the two cases are the same. 
We have calculated the electron density profiles of the ripple phase of 
{\sl l} 
and {\sl dl}-DMPC bilayers at various temperatures and a 
relative humidity of 98${\pm}$2${\%}$. 
The profiles at 
24 $^{o}$C are presented in Figs. 1 and 2. The structural parameters
of the ripple phase 
at two different temperatures are given in Table II. As one can observe, 
the ripples formed by chiral and racemic DMPC bilayers do not differ 
appreciably. The peak-to-peak amplitude in both cases is approximately 18 {\AA}, 
and the ratio of the length of the major arm to that of the minor arm
is found 
to be 2.1 and 2.4 for {\sl l}-DMPC and {\sl dl}-DMPC bilayers, 
respectively. Furthermore, the thickness of the bilayer in the 
two arms is different in both cases. This can be explained in terms of 
an average tilt of the hydrocarbon chains along the ripple wavevector [13].
The minor differences in the two figures arise most probably because we have 
not been able to apply  
absoption corrections. 
Thus at all temperatures studied, the structure of the ripple phase in
chiral and racemic DMPC bilayers is practically identical.
\\
\noindent
{\bf V. CONCLUSION}
\indent

In this brief report we have presented electron density profiles of 
chiral and racemic DMPC bilayers in the ripple phase. The maps show
asymmetric ripples in both the systems. Moreover, all 
the structural features in the two cases are found to be practically
identical 
under similar experimental conditions. The present study thus unambiguously
confirms our earlier conclusion that the structure of the {\sl P$_{\beta^{\prime}}$}
phase does 
not depend on the chirality of the lipid molecules constituting the
bilayer [11].
\\
{\bf Acknowledgements:} We thank Yashodhan Hatwalne and K. Usha for many
helpful discussions and HKL Research Inc. for the use of their software.
\\

{\bf References:}
\indent
\begin{enumerate}

\item A. Tardieu, V. Luzzati, and F. C. Reman, J. Mol. Biol. {\bf75,} 711 
(1973).

\item F. C. MacKintosh, Current Opinion in Colloid and Interface Science
{\bf 2,} 382 (1997).

\item M. J. Janiak, D. M. Small, and G. G. Shipley, J. Biol. Chem. 
{\bf 254,} 6068 (1979).

\item G. S. Smith, E. B. Sirota, C. R. Safinya, and N. A. Clark, Phys. Rev. 
Lett. {\bf 60,} 813 (1988).

\item  D. C. Wack and W. W. Webb, Phys. Rev. A {\bf 40,} 2712 (1989).

\item M. P. Hentschel and  F. Rustichelli, Phys. Rev. Lett. {\bf 66,} 903 
(1991).

\item J. T. Woodward and J. A. Zasadzinski, Biophys. J. {\bf 72,} 964 
(1997).

\item T. C. Lubensky and F. C. MacKintosh, Phys. Rev. Lett. {\bf 71,} 1565 
(1993).

\item C-.M. Chen, T.C. Lubensky and F.C. MacKintosh, Phys. Rev. E {\bf 51},
504 (1995).

\item Symmetric (asymmetric) ripples correspond to a rectangular (oblique) 
  unit cell.

\item J. Katsaras and V. A. Raghunathan, Phys. Rev. Lett. {\bf 74,} 2022 
(1995).

\item J. Katsaras, R. F. Epand, and R. M. Epand, Phys. Rev. E {\bf 55,} 
3751 (1997).

\item K. Sengupta, V. A. Raghunathan and J. Katsaras (to be published).

\item W.-J. Sun, S. Tristram-Nagle, R. M. Suter, and J. F. Nagle, 
Proc. Natl. Acad. Sci. USA {\bf 93,} 7008 (1996).

\end{enumerate}

\centerline {\bf Table I.}

\centerline {The observed and calculated stucture factors at 24$^o$C and 98$\%$ RH}
\hrulefill
\begin{tabbing}
000000 \= 000000 \= 00000000 \= 000000000 \= 000000000    \= 00000 \=000000000
\= 0000000000 \= 0000000000 \kill
\>           \>        \>   l-DMPC   \>       \>                \>  dl-DMPC  \> \\ 
\>    {\bf h}\> {\bf k}\> $|${\bf F$_{o}$}$|$\> {\bf F$_{c}$}\> \> $|${\bf F$_{o
}$}$|$\> {\bf F$_{c}$} \\
\>       1 \>       0  \>       100.0 \>   -80.2 \>\> 100.0 \> -117.1\\
\>1 \>-1 \>42.5 \>-38.0 \> \>  59.2 \> -32.9\\
\>2 \>0  \>29.7 \>-19.8 \> \> 65.0 \> -43.3\\
\>2 \>-1 \>43.6 \>-40.5 \> \> 68.0 \> -55.2\\
\>2 \>-2 \>9.3 \>-15.4 \>\>  29.0 \> -38.5\\
\>2 \>2  \>15.7 \>-14.6 \> \> 21.7 \> -3.2\\
\>2 \>3  \>10.6 \>7.9 \> \> 10.6 \> 0.3\\
\>3 \>0  \>21.2 \>-2.9 \> \> 32.9 \> 11.5\\
\>3 \>-1 \> 42.4 \>33.9 \> \> 53.3 \> 39.3\\
\>3 \>-2 \> 29.2 \>32.0 \> \> 35.0 \> 38.3\\
\>3 \>2. \>9.5 \>17.1 \> \> 9.0 \> 16.0\\
\>3 \>3. \>16.5 \>-15.9 \> \> 12.3\>  -0.5\\
\>3 \>4  \>16.3 \>10.3 \> \> 13.3 \> 2.0\\
\>3 \>5  \>12.4 \>-4.2 \> \> - \> -\\
\>4 \>0  \>29.9 \>27.9 \> \> 31.1 \> 22.5\\
\>4 \>-1 \> 25.5 \>-31.2 \> \>  33.1 \> -37.4\\
\>4 \>-2 \>55.5 \>-79.5 \> \> 56.8 \> -57.5\\
\>4 \>-3 \>27.7 \>-33.9 \> \> 27.9 \> -35.8\\
\>5 \>0  \>8.5 \>0.6 \> \>  9.6 \> 4.6\\
\>5 \>-1 \> 5.9 \>0.8 \> \> 12.0 \> -0.3\\
\>5 \>-2 \> 7.8 \>-10.1  \>\> 10.8 \> -29.4\\
\>6 \>0  \>4.4 \>1.2 \> \> 6.4 \> -5.6\\
\>6 \>-1 \> 5.9 \>-5.4  \>\> - \> -\\
\>7 \>0 \> -    \>-    \> \> 5.5 \> -2.2 \\
\>9 \>0 \>-     \>-     \> \> 4.5 \> -1.3 \\
\end{tabbing}
\hrulefill \\

\centerline {\bf Table II.} 

\centerline {The structural parameters of the ripple phase} 

\hrulefill \\
\begin{tabbing}
000\= DL-dmpc000 \=  1000.0000 \= 100000000 \= 100000000 \= 100000000 \= 100000
00 \=100000000 \=
 10000000 \= 100000000 \kill
\> \> \> { $\gamma$($^o$)} \> { $\lambda$}({\AA}) \>
{ D} \> { $\lambda_{1}$}(\AA) \> { A}(\AA) \\ \\
  \> l-dmpc\>  24 $^o$ C \> 99$\pm$1 \> 142$\pm$2 \> 56$\pm$1 \>97$\pm$2 \>

18$\pm$1 \\
 \>    \> 21 $^o$ C \> 99 \> 145 \> 56 \>98 \>18 \\ \\
  \>dl-dmpc \> 24 $^o$ C \> 98 \> 141 \> 56 \>100 \>19 \\
\>  \> 21 $^o$ C \> 98 \> 140 \> 56 \>99 \>18 \\

\end{tabbing}
\hrulefill \\

\newpage
\begin{figure} [c]
\hbox{\centerline{\psfig{figure=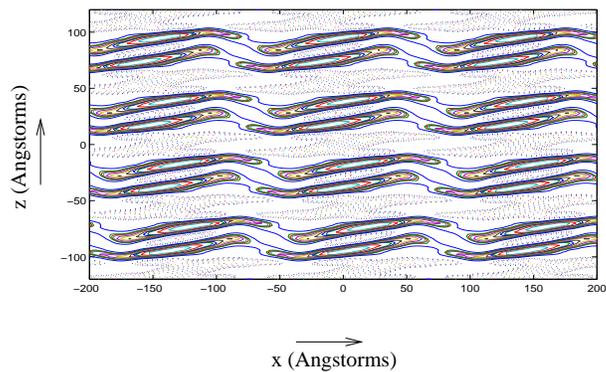,width=8cm}}}
\caption{Electron Density Map of the ripple phase of dl-DMPC at 24$^o$C and 98$
\%$ RH.
The positive (negative) contours are represented by solid (dotted) lines.
The regions with positive electron density correspond to the head groups.
Note that the thickness of the bilayers is about 40 {\AA}, whereas that of the
water regi
on is
about 15 {\AA}}
\end{figure}

\begin{figure} [c]
\hbox{\centerline{\psfig{figure=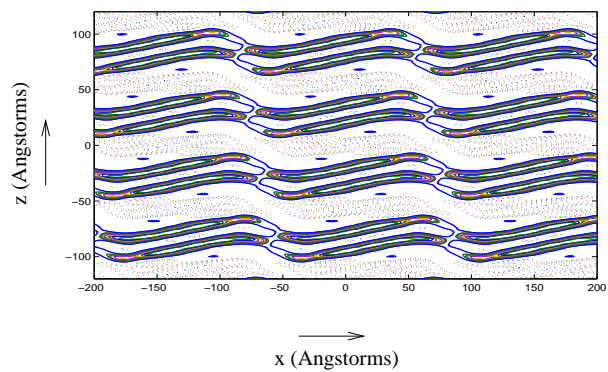,width=8cm}}}
\caption{Electron Density Map of ripple phase of l-DMPC at 24$^o$C and 98$\%$ R
H.}
\end{figure}

\end{document}